\documentclass[12pt,draftcls,journal,leterpaper,twosides,onecolumn]{IEEEtran}
\usepackage[nolists,nomarkers,noheads]{endfloat}%figs in the end
\usepackage{amsmath,amssymb,amscd,latexsym,dsfont}
\usepackage{float,color,graphicx,subfigure}
\usepackage{multirow,multicol}
\usepackage{comment,cite}
\usepackage{enumerate}
\usepackage{stfloats}
\usepackage{psfrag}
\usepackage{cite}

\title{On BICM receivers for TCM transmission}
\author{%
\IEEEauthorblockN{Alex Alvarado, Leszek Szczecinski\IEEEauthorrefmark{4}, and Erik Agrell\\}
\IEEEauthorblockA{Department of Signals and Systems, Communication Systems Group\\ Chalmers University of Technology, Gothenburg, Sweden\\}
\IEEEauthorblockA{\IEEEauthorrefmark{4}INRS-EMT, Montreal, Canada\\}

\emph{alex.alvarado@chalmers.se, leszek@emt.inrs.ca, agrell@chalmers.se}

\thanks{This work was partially supported by the European Commission under projects NEWCOM++ (216715) and FP7/2007-2013 (236068), and by the Swedish Research Council, Sweden (2006-5599). When this work was submitted for publication, L.~Szczecinski was on sabbatical leave with CNRS, Laboratory of Signals and Systems, Gif-sur-Yvette, France. Parts of this work will be presented at the Allerton Conference on Communication, Control, and Computing, IL, USA, September 2010.}
}%

\newcommand{\tr}[1]{\textrm{#1}}
\newcommand{\mb}[1]{\mathbf{#1}}
\newcommand{\ov}[1]{\overline{#1}}

\newcommand{\mc}[1]{\mathcal{#1}}

\newcommand{\set}[1]{\{#1\}}

\newcommand{\cd}{\cdot}
\newcommand{\ld}{\ldots}

\newcommand{\nchoosek}[2]{{{#1} \choose {#2}}}

\newtheorem{theorem}{Theorem}
\newtheorem{example}{Example}
\newtheorem{corollary}[theorem]{Corollary}
\newtheorem{definition}{Definition}

\newcommand{\ie}{i.e.,~}
\newcommand{\eg}{e.g.,~}

\newcommand{\cf}{cf.~}
\newcommand{\mat}[1]{\mathbf{#1}}
\newcommand{\bb}{\boldsymbol{b}}
\newcommand{\be}{\boldsymbol{e}}
\newcommand{\bs}{\boldsymbol{s}}
\newcommand{\bi}{\boldsymbol{i}}

\newcommand{\bc}{\boldsymbol{c}}
\newcommand{\bx}{\boldsymbol{x}}
\newcommand{\bz}{\boldsymbol{z}}
\newcommand{\by}{\boldsymbol{y}}
\newcommand{\matC}{\mat{C}}
\newcommand{\matE}{\mat{E}}
\newcommand{\matS}{\mat{S}}
\newcommand{\matL}{\mat{L}}
\newcommand{\matzero}{\mat{0}}

\begin{document}

\maketitle

\begin{abstract}
Recent results have shown that the performance of bit-interleaved coded modulation (BICM) using convolutional codes in nonfading channels can be significantly improved when the interleaver takes a trivial form (BICM-T), \ie when it does not interleave the bits at all. In this paper, we give a formal explanation for these results and show that BICM-T is in fact the combination of a TCM transmitter and a BICM receiver. To predict the performance of BICM-T, a new type of distance spectrum for convolutional codes is introduced, analytical bounds based on this spectrum are developed, and asymptotic approximations are also presented. It is shown that the minimum distance of the code is not the relevant optimization criterion for BICM-T. Optimal convolutional codes for different constrain lengths are tabulated and asymptotic gains of about 2~dB are obtained. These gains are found to be the same as those obtained by Ungerboeck's one-dimensional trellis coded modulation (1D-TCM), and therefore, in nonfading channels, BICM-T is shown to be asymptotically as good as 1D-TCM.
\end{abstract}


%%%%%%%%%%%%%%%%%%%%%%%
\section{Introduction}\label{Sec:Introduction}

Coded modulation (CM) was introduced in 1974 when Massey proposed the idea of jointly designing the channel encoder and modulator \cite{Massey74}. This inspired Ungerboeck's trellis coded modulation (TCM) \cite{Ungerboeck82}, and Imai and Hirakawa's multilevel coding \cite{Imai77}. Bit-interleaved coded modulation (BICM) \cite{Zehavi92,Caire98,Fabregas08_Book} appeared in 1992 as an alternative for CM in fading channels. One particularly appealing feature of BICM is that all the operations are bit-wise, \ie off-the-shelf binary codes and Gray-mapped constellations are used at the transmitter's side and connected via a bit-level interleaver. At the receiver's side, reliability metrics for the coded bits (L-values) are calculated by the demapper, de-interleaved, and then fed to a binary decoder. This structure gives the designer the flexibility to choose the modulator and the encoder independently, which in turn allows, for example, for an easy adaptation of the transmission to the channel conditions (adaptive modulation and coding). This flexibility is arguably the main advantage of BICM over other CM schemes, and also the reason of why it is used in almost all of the current wireless communications standards, \eg HSPA, IEEE 802.11a/g, IEEE 802.16, and DVB-S2 \cite[Ch.~1]{Fabregas08_Book}.

Bit-interleaving before modulation was introduced in Zehavi's original paper \cite{Zehavi92} on BICM. Bit-interleaving is indeed crucial in fading channels since it guarantees that consecutive coded bits to be sent over symbols affected by independent fades. This results in an increase (compared to TCM) of the so-called code diversity (the suitable performance measure in fading channels), and therefore, BICM is the preferred alternative for CM in fading channels. BICM can also be used in nonfading channels. However, in this scenario, and compared with TCM, BICM gives a smaller minimum Euclidean distance (the proper performance metric in nonfading channels), and also a smaller constraint capacity \cite{Caire98}. If a Gray labeling is used, the capacity loss is small, and therefore, BICM is still considered valid option for CM over nonfading channels. However, the decrease in minimum Euclidean distance makes BICM less appealing than TCM in nonfading channels. 

The use of a bit-level interleaver in nonfading channels has been inherited from the original works on BICM by Zehavi \cite{Zehavi92} and Caire~\emph{et al.} \cite{Caire98}. It simplifies the performance analysis of BICM and is implicitly considered mandatory in the literature. However, the reasons for its presence are seldom discussed. 

Previously, we have shown in \cite{Alvarado09c} how---by using multiple interleavers---the performance of BICM can be improved in nonfading channels. Recently, however, it has been shown in \cite{Stierstorfer10} that in nonfading channels, considerably larger gains (a few decibels) can be obtained if the interleaver is \emph{completely removed} from the tranceiver's configurations. In other words, it was shown that in nonfading channels BICM without an interleaver performs better than the conventional configurations of \cite{Zehavi92,Caire98}. The results presented in \cite{Stierstorfer10} are solely numerical and an explanation behind such an improvement is not given. In particular, \cite{Stierstorfer10} does not explain why the obtained gains depend on the constraint length of the convolutional code (CC). Nevertheless, in \cite{Stierstorfer10} some intuitive explanations (using the notion of unequal error protection) and a bit labeling optimization are presented. 

In this paper, we present a formal study of BICM with trivial interleavers (BICM-T) in nonfading channels, \ie the BICM system introduced in \cite{Stierstorfer10} where no interleaving is performed. We recognize BICM-T as the combination of a TCM transmitter and a BICM receiver and we develop analytical bounds that give a formal explanation of why BICM-T with CCs performs well in nonfading channels. We also introduce a new type of distance spectrum for the CCs which allows us to analytically corroborate the results presented in \cite{Stierstorfer10}. These gains are shown to appear even for one of the simplest configuration one could think of, \ie when the constraint length $K=3$ convolutional code with generators $(5,7)$ is used together with 4-ary pulse amplitude modulation (PAM). Asymptotic bounds are also developed and used to show that for the $(5,7)$ code and 4-PAM, an asymptotic gain of 2.55~dB is obtained compared to an uncoded system with the same spectral efficiency. Motivated by the fact that this gain is the same obtained by Ungeroboeck's one-dimensional TCM (1D-TCM), we search and tabulate optimum convolutional codes for BICM-T. We show that a properly design BICM system without interleaving performs asymptotically as well as 1D-TCM, and therefore, BICM-T should be considered as a good alternative for CM in nonfading channels. The main contribution if this paper is to present an analytical model for BICM-T which is used to explain the results presented in \cite{Stierstorfer10} and also to design a BICM-T system in nonfading channels.

%%%%%%%%%%%%%%%%%%%%%%%
\section{System Model and Preliminaries}\label{Sec:Model}

Throughout this paper, we use boldface letters $\bc_t=[c_{1,t},\ld,c_{L,t}]$ to denote lenght-$L$ row vectors and capital boldface letters $\matC=[\bc_1^\tr{T},\ld, \bc_N^\tr{T}]$ to denote matrices, where $(\cd)^\tr{T}$ denotes transposition. We use $d_\tr{H}(\matC)$ to denote the total Hamming weight of the matrix $\matC$. We denote probability by $\tr{Pr}(\cd)$ and the probability density function (pdf) of a random variable $\Lambda$ by $\tr{p}_\Lambda(\lambda)$. The convolution between two pdfs is denoted by $\tr{p}_{\Lambda_1}(\lambda)*\tr{p}_{\Lambda_2}(\lambda)$ and $\set{\tr{p}_\Lambda(\lambda)}^{*w}$ denotes the $w$-fold self-convolution of the pdf $\tr{p}_\Lambda(\lambda)$. A Gaussian distribution  with mean value $\mu$ and variance $\sigma^2$ is denoted by $\mc{N}(\mu,\sigma^2)$, the Gaussian function with the same parameters by $\psi(\lambda;\mu,\sigma)\triangleq\frac{1}{\sqrt{2\pi}\sigma}\exp(-\frac{(\lambda-\mu)^{2}}{2\sigma^{2}})$, and the Q-function by $Q(x) \triangleq\frac{1}{\sqrt{2\pi}}\int_{x}^{\infty}\exp{\left(-\frac{u^2}{2}\right)}\,du$. All the polynomial generators of the convolutional codes (CC) are given in octal notation.

\subsection{System Model}\label{Sec:Model:Model}

\begin{figure}[t]
\psfrag{b1}[cB][cB][0.85]{$\bi$}%
\psfrag{enc}[cB][cB][0.9]{ENC}%
\psfrag{Ct}[cB][cB][0.9]{$\matC$}%
\psfrag{CtPi}[cB][cB][0.9]{$\Pi(\matC)$}%
\psfrag{pi}[cB][cB][0.9]{$\Pi$}%
\psfrag{M}[cB][cB][0.9]{$\Phi$}%
\psfrag{s}[cB][cB][0.9]{$\bx$}%
\psfrag{z}[cB][cB][0.9]{$\bz$}%
\psfrag{r}[cB][cB][0.9]{$\by$}%
\psfrag{M1}[cB][cB][0.9]{$\Phi^{-1}$}%
\psfrag{pi2}[cB][cB][0.9]{$\Pi^{-1}$}%
\psfrag{LPi}[cB][cB][0.9]{$\Pi(\matL)$}
\psfrag{L}[cB][cB][0.9]{$\matL$}
\psfrag{dec}[cB][cB][0.9]{DEC}%
\psfrag{b21}[cB][cB][0.9]{$\hat{\bi}$}%
\begin{center}
    \includegraphics[width=0.95\columnwidth]{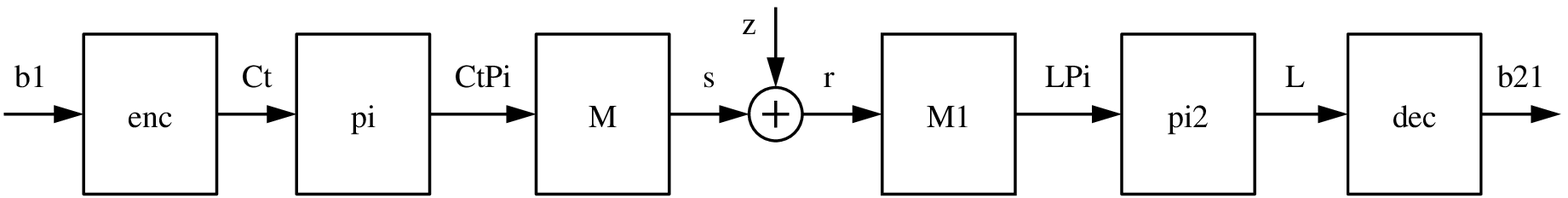}
    \caption{Model of BICM transmission.}
    \label{Sec:Preliminaries:system_model}
\end{center}
\end{figure}

The BICM system model under consideration is presented in Fig.~\ref{Sec:Preliminaries:system_model}. We use a constraint length $K$ rate $R=\frac{1}{2}$ convolutional encoder connected to a 16-ary quadrature amplitude modulation (16-QAM) labeled by the binary reflected Gray code (BRGC)\cite{Agrell04}. This configuration is indeed very simple yet practical yielding a spectral efficiency of two bits per real channel use. This example is not restrictive, of course, yet simplifies the presentation of the main ideas. The generalization to other modulations and coding rate is naturally possible but would obviously increase the complexity of notation potentially hindering the main concepts of the analysis presented in this paper. 

The input sequence $\bi=[i_{1},\ld, i_{N}]$ is fed to the encoder (ENC) which at each time instant $t=1,\ld,N$ generates two coded bits $\bc_{t} =[c_{1,t},c_{2,t}]$. We use the matrix $\matC=[\mb{c}^\tr{T}_{1},\ld,\mb{c}^\tr{T}_{N}]$ of size $2\times N$ to represent the transmitted codeword. These coded bits are interleaved by $\Pi$, where the different interleaving alternatives will be discussed in detail in Sec.~\ref{Sec:Model:Int}. The coded and interleaved bits are then mapped to a 16-QAM symbol, where the 16-QAM constellation is formed by the direct product of two 4-ary pulse amplitude modulation (4-PAM) constellations labeled by the BRGC. Therefore, we analyze the real part of the constellation only, \ie one of the constituent 4-PAM constellations. The mapper is defined as $\Phi: \set{[1 1],[1 0], [0 0], [0 1]} \rightarrow \set{-3\Delta, -\Delta, \Delta, 3\Delta}$, where we define
\begin{align}\label{Delta}
\Delta\triangleq\frac{1}{\sqrt{5}}
\end{align}
so that the PAM constellation normalized to unit average symbol energy, \ie $E_\tr{s}=1$.

A quick inspection of the BRGC for 4-PAM reveals that the BRGC offers unequal error protection (UEP) to the transmitted bits depending on their position. In particular, the bit at the first position ($k=1$) receives higher protection\footnote{The ``protection'' may be defined in different ways, where probably the simplest one is the bit error probability per bit position at the demapper's output.} than the bit at the second position $k=2$. More details about this can be found in \cite{Alvarado09c}. Moreover, for $k=2$ a bit labeled by zero (inner constellation points) will receive a lower protection than a bit labeled by one transmitted in the same bit position (outer constellation points), and therefore, the binary-input soft-output (BISO) channel for $k=2$ is nonsymmetric. To simplify the analysis, we ``symmetrize'' the channel by randomly inverting the bits before mapping them to the 4-PAM symbol, \ie $\tilde{\matC}=\Pi(\matC) \oplus \matS$, where $\oplus$ represents modulo-2 element-wise addition and the elements of the matrix $\matS=[\bs_1^\tr{T},\ld,\bs_N^\tr{T}]\in\set{0,1}^{2\times N}$ where $\bs_t=[s_{1,t},s_{2,t}]$ are randomly generated vectors of bits. Such a scrambling symmetrizes the BISO channel but it does not eliminate the UEP. We note that the scrambling is introduced only to simplify the analysis, and therefore, it is not shown in Fig.~\ref{Sec:Preliminaries:system_model} nor used in the simulations. This symmetrization was in fact proposed in \cite{Caire98}, and as we will see in Sec.~\ref{Sec:DiscAndApp}, the bounds developed based on this symmetrization perfectly match the numerical simulations.

At each time $t=1,\ld,N$, the coded and scrambled bits $\tilde{\bc}_{t}$ are mapped to a symbol $x_{t}$, where $x_{t}=\Phi(\tilde{\bc}_{t})\in \mc{X}$ and $\mc{X}$ is the 4-PAM constellation. The symbols $x_{t}$ are sent over an additive white Gaussian noise (AWGN) channel so the received signal is given by $y_{t}=x_{t}+z_{t}$, where $z_{t}$ is a zero-mean Gaussian noise with variance $N_{0}/2$. The signal-to-noise ratio is defined as $\gamma\triangleq E_\tr{s}/N_0=1/N_{0}$.  At the receiver's side, reliability metrics for the bits are calculated by the demapper $\Phi^{-1}$ in the form of logarithmic-likelihood ratios (L-values) as
\begin{align}\label{L-value.1}
  \tilde{l}_{k,t}&=\log\frac{\tr{Pr}(\tilde{c}_{k,t}=1 | y_{t})}{\tr{Pr}(\tilde{c}_{k,t}=0 | y_{t})}.
\end{align}
Since $\tilde{c}_{k,t}=c_{k,t} \oplus s_{k,t}$, it can be shown that $l_{k,t}=(-1)^{s_{k,t}} \tilde{l}_{k,t}$, \ie after ``descrambling'', the sign of the L-values is changed using $(-1)^{s_{k,t}}$. These L-values are deinterleaved and then passed to the decoder which calculates an estimate of the information sequence $\hat{\bi}$.

\subsection{The interleaver}\label{Sec:Model:Int}

Throughout this paper, three interleaving alternatives will be analyzed, \cf the block $\Pi$ in Fig.~\ref{Sec:Preliminaries:system_model}. The first interleaving alternative is BICM with a single interleaver (BICM-S) introduced in \cite{Caire98}. It is the most commonly used in the literature and corresponds to an interleaver that randomly permutes the bits $\matC$ prior to modulation, where the permutation is random in two ``dimensions,'' \ie it permutes the bits over the bit positions and over time. The second alternative is BICM with multiple interleavers (M-interleavers, BICM-M) where the interleaver permutes the bits randomly only over time (and not over the bit positions). This can be seen as a particularization of the interleaver of BICM-S following an additional constraint: bits from the $k$th encoder's output must be assigned to the $k$th modulator's input. BICM-M was formally analyzed in \cite{Alvarado09c} and in fact corresponds to the original model introduced by Zehavi in \cite{Zehavi92} (BICM) and Li in \cite{Li98} (BICM with iterative decoding, BICM-ID). Recently, M-interleavers have also been proven to be asymptotically optimum for BICM-ID \cite{Alvarado10b}. The last interleaving alternative, on which this paper focuses, is BICM with a trivial interleaver (BICM-T), \ie when the interleaver $\Pi$ in Fig.~\ref{Sec:Preliminaries:system_model} is simply not present \cite{Stierstorfer10}.  

\begin{figure}[t]
\psfrag{b1}[cB][cB][1]{$i_t$}%
\psfrag{enc}[cc][cc][1]{ENC}%
\psfrag{c1t}[cB][cB][1]{$c_{1,t}$}%
\psfrag{c2t}[cB][cB][1]{$c_{2,t}$}%
\psfrag{M}[cB][cB][1]{$\Phi$}%
\psfrag{s}[cB][cB][1]{$x_t$}%
\psfrag{z}[cB][cB][1]{$z_t$}%
\psfrag{r}[cB][cB][1]{$y_t$}%
\psfrag{M1}[cB][cB][1]{$\Phi^{-1}$}%
\psfrag{l1t}[cB][cB][1]{$l_{1,t}$}
\psfrag{l2t}[cB][cB][1]{$l_{2,t}$}
\psfrag{dec}[cB][cB][1]{DEC}%
\psfrag{b21}[cB][cB][1]{$\hat{i}_t$}%
\begin{center}
    \includegraphics[width=0.78\columnwidth]{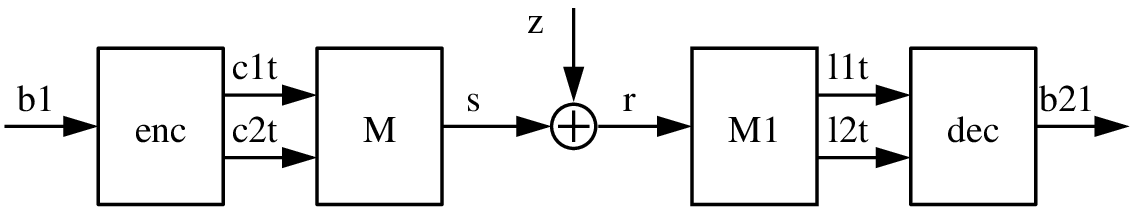}
    \caption{BICM-T system analyzed in this paper for any time instant $t$.}
    \label{Sec:Preliminaries:BICM-T}
\end{center}
\end{figure}

When BICM-T is considered, the resulting system is the one shown in Fig.~\ref{Sec:Preliminaries:BICM-T}. A careful examination of Fig.~\ref{Sec:Preliminaries:BICM-T} reveals that the structure of the transmitter of BICM-T is the same as the transmitter of Ungerboeck's one-dimensional TCM \cite{Ungerboeck82} or the TCM transmitter in \cite[Fig.~4.17]{Jamali94_Book}. The transmitter of BICM-T can also be considered a particular case of the so-caled ``general TCM'' \cite[Fig.~18.11]{Lin03_Book} when  $k=\tilde{k}$ (using the notation of \cite{Lin03_Book}) and when the BRGC is used instead of Ungerboeck's set-partitioning. The receiver of BICM-T in Fig.~\ref{Sec:Preliminaries:BICM-T} corresponds to a conventional BICM receiver, where L-values for each bit are computed and fed to a soft-input Viterbi decoder (VD). The difference between this receiver's structure and a TCM receiver is that bit-level processing is used instead of a symbol-by-symbol VD. In conclusion, the BICM-T system introduced in \cite{Stierstorfer10} is simply a BRGC-based TCM transmitter used in conjunction with a BICM receiver. Nevertheless, through this paper, we use the name BICM-T to reflect the fact that this transmitter/receiver structure can be considered as a particular case of the BICM system in \cite{Zehavi92,Caire98}, where the interleaver takes a trivial form.

\subsection{The Decoder}\label{Sec:Model:Dec}

A maximum likelihood sequence decoder (\eg the VD) chooses the most likely coded sequence $\hat{\matC}$ using the vector of channel observations $\mb{y}=[y_{1},\ld,y_{N}]$ as
\begin{align}\label{decision_DEC}
 \hat{\matC}	& =\max_{\matC \in \mc{D}} \left\{ \log\bigl(\tr{Pr}\set{ {\matC} | \mb{y} } \bigr) \right\}\\
 			& = \max_{\matC \in \mc{D}} \left\{ \log\left(\prod_{t=1}^{N}\tr{Pr}\set{ \bc_{t} | y_t }\right) \right\}\label{decision_DEC2},
\end{align}
where $\mc{D}$ is the set of all codewords, where to pass from \eqref{decision_DEC} to \eqref{decision_DEC2} we used the memoryless property of the channel. If we assume that the bits $[c_{1,t},c_{2,t}]$ are independent, we obtain
\begin{align}
\log\left(\prod_{t=1}^{N}\tr{Pr}\set{ \bc_{t} | y_t }\right)	& = \log\left(\prod_{k=1}^{2}\prod_{t=1}^{N}\tr{Pr}\set{ c_{k,t} | \mb{y} }\right).\label{Pc}
\end{align}

Under this independence assumption and by using the relation between an L-value $l$ and the bit's probabilities of being $b\in\set{0,1}$
\begin{align}\label{Pby}
\tr{Pr}\set{b|y}=\frac{\tr{e}^{bl}}{1+\tr{e}^{l}},
\end{align}
we obtain
\begin{align}
\log\left(\prod_{k=1}^{2}\prod_{t=1}^{N}\tr{Pr}\set{ c_{k,t} | \mb{y} }\right) & =\sum_{k=1}^{2}\sum_{t=1}^{N}\log\bigl(\tr{Pr}\set{ c_{k,t} | \mb{y} }\bigr)\nonumber\\
			& = \sum_{k=1}^{2}\sum_{t=1}^{N} c_{k,t} l_{k,t} - \sum_{k=1}^{2}\sum_{t=1}^{N}\log(1+\exp(l_{k,t}))\label{Pc2}.
\end{align}

Since the second term in \eqref{Pc2} is independent of $\matC$, it is irrelevant to the decision of the decoder in \eqref{decision_DEC2}. Therefore, the final decision of the decoder can be written as
\begin{align}\label{Pc3}
 \hat{\matC}=\max_{\matC \in \mc{D}} \left\{ \sum_{k=1}^{2}\sum_{t=1}^{N} c_{k,t} l_{k,t} \right\}.
\end{align}

In a BICM system with convolutional codes, the decoder is implemented using an off-the-shelf soft-input VD, which assumes that the bits are independent, and thus, uses the relation in \eqref{Pc} (\ie it uses the decision rule in \eqref{Pc3}). The relation in \eqref{Pc} is in indeed valid when BICM-S \cite{Caire98} or BICM-M \cite{Zehavi92,Alvarado09c,Alvarado10b} configurations are used, since in those cases, the use of a random interleaver (\cf Sec.~\ref{Sec:Model:Int}) assure that the bits $[c_{1,t},c_{2,t}]$ are transmitted in different symbols, and therefore, are affected by different noise realizations. 

However, when BICM-T with a soft-input VD is considered, and since the bits $[c_{1,t},c_{2,t}]$ are affected by the same noise realization, the relation in \eqref{Pc} does not hold, \ie the two L-values passed to the decoder at any time instant $t$ are not independent. Nevertheless, the decoder treats the bits as independent and still uses the decision rule in \eqref{Pc3}. In principle, it would be possible to design a decoder for BICM-T that takes into account this inconsistency, \ie a decoder that does not assume independent bits. However, this is out of the scope of this paper and would also go against the flexibility offered by BICM. Moreover, we will show in the following section that even with this inconsistency, BICM-T in nonfading channels outperforms BICM-S and BICM-M.

\section{Performance Evaluation}\label{Sec:Performance}

\subsection{BER Performance}\label{Sec:Performance:BER}

Because of the symmetrization of the channel, we can, without loss of generality, assume that the all-zero codeword was transmitted. We define $\mc{E}$ as the set of codewords corresponding to paths in the trellis of the code diverging from the zero-state at the arbitrarily chosen instant $t=t_0$, and remerging with it after $T$ trellis stages. We also denote these codewords as $\matE\triangleq [\be_1^\tr{T},\ld, \be_T^\tr{T}]$, where $\be_t=[e_{1,t},e_{2,t}]$. Then, the bit error rate (BER) can be upper-bounded using a union bound (UB) as
\begin{align}
  \tr{BER} \leq \tr{UB}&\triangleq \sum_{\matE\in\mc{E}} \tr{PEP}(\matE)d_\tr{H}(\bi_\matE),\label{UB}
\end{align}
where $d_{\tr{H}}(\bi_\matE)$ is the Hamming weight of the input sequence $\bi_\matE$ corresponding to the codeword $\matE$, and the pairwise error probability (PEP) is given by (\cf\eqref{Pc3})
\begin{align}
  \tr{PEP}(\matE)	& =\tr{Pr}\left\{ \sum_{t=t_0}^{t_0+T-1}\bigl( e_{1,t} l_{1,t}+ e_{2,t} l_{2,t}\bigr) > 0 \right\}\label{PEP3}.
\end{align}
The general expression for the PEP in \eqref{PEP3} and the UB in \eqref{UB} reduce to well-known particular cases  if simplifying assumptions for the distribution of $l_{k,t}$ are adopted. 

\subsubsection{Independent and identically distributed L-values (BICM-S)}

In BICM-S, the L-values $l_{k,t}$ passed to the decoder are independent and identically distributed (i.i.d.). They can be described using the conditional pdf $\tr{p}(\lambda|b)$ with $b\in\set{0,1}$ and where the pdf is independent of $k$ and $t$. In this case, the PEP in \eqref{PEP3} depends only on the Hamming weight of the codeword $\matE$, \ie
\begin{align}
   \tr{PEP}(\matE) 	& = \tr{PEP}_\tr{S}( d_\tr{H}(\matE) )\nonumber\\
   				& =\int_{0}^{\infty} \set{\tr{p}(\lambda|b=0)}^{* d_\tr{H}(\matE)} \, \tr{d}\lambda.\label{PEP.iid}
\end{align}
The UB in \eqref{UB} can be expressed as
\begin{align}
\tr{UB}_\tr{S}	&=\sum_{w} \tr{PEP}_\tr{S}(w) \sum_{\matE\in\mc{D}_w} d_\tr{H}(\bi_\matE)\label{UB.iid1}\\
 		&=\sum_{w} \tr{PEP}_\tr{S}(w) \beta_{w}^\mc{C},\label{UB.iid2}
\end{align}
where $\mc{D}_w$ represents the set of codewords with Hamming weight $w$, \ie $\mc{D}_w\triangleq \set{\matE\in\mc{E}:d_\tr{H}(\matE)=w}$. To pass from \eqref{UB.iid1} to \eqref{UB.iid2} we group the codewords $\matE$ that have the same Hamming weight and add their contributions, which results in the well-known weight distribution spectrum of the code $\beta_{w}^\mc{C}$. The expression in \eqref{UB.iid2} is the most common expression for the UB for BICM, \cf\cite[eq.~(26)]{Caire98},~\cite[eq.~(4.12)]{Fabregas08_Book}.

\subsubsection{Independent but not identically distributed L-values (BICM-M)}

In BICM-M, the L-values passed to the each decoder's input are independent, however, their conditional pdf depends on the bit's position $k=1, 2$. Thus, the L-values are  modeled by the set of conditional pdfs $\set{\tr{p}_{1}(\lambda|b),\tr{p}_{2}(\lambda|b)}$. The PEP in this case is given by 
\begin{align}
   \tr{PEP}(\matE)	& = \tr{PEP}_\tr{M}(\ov{w}_{\matE,1},\ov{w}_{\matE,2})\nonumber\\
   				& = \int_{0}^{\infty} \set{\tr{p}_{1}(\lambda|b_1=0)}^{* \ov{w}_{\matE,1}} * \set{\tr{p}_{2}(\lambda|b_2=0)}^{* \ov{w}_{\matE,2}}\,\tr{d}\lambda,\label{PEP.inid}
\end{align}
where $\ov{w}_{\matE,k}$ is the Hamming weight of the $k$th row of $\matE$. The UB in \eqref{UB} can be expressed as
\begin{align}
\tr{UB}_\tr{M} 	&=\sum_{w_{1},w_{2}} \tr{PEP}_\tr{M}(w_{1},w_{2}) 
  \sum_{\matE\in\mc{D}_{w_1,w_2}} d_\tr{H}(\bi_\matE) \nonumber\\
 &=\sum_{w_{1},w_{2}} \tr{PEP}_\tr{M}(w_{1},w_{2}) \beta_{w_{1},w_{2}}^\mc{C},\label{UB.inid2}
\end{align}
where $\mc{D}_{w_1,w_2}$ is the set of codewords with \emph{generalized} Hamming weight $[w_1,w_2]$ ($w_k$ in its $k$th row), \ie $\mc{D}_{w_1,w_2}\triangleq \set{\matE\in \mc{E}:w_{1}=\ov{w}_{\matE,1},w_{2}=\ov{w}_{\matE,2}}$,  and $\beta_{w_{1},w_{2}}^\mc{C}$ is the generalized weight distribution spectrum of the code that takes into account  the errors at each encoder's output separately. The UB in \eqref{UB.inid2} was shown in \cite{Alvarado09c} to be useful when analyzing the UEP introduced by the binary labeling and also to optimize the interleaver and the code.

\subsubsection{BICM without bit-interleaving (BICM-T)}

For BICM-T, yet a different particularization of \eqref{PEP3} must be adopted. Let $\Lambda_\matE$ be the metric associated to the codeword $\matE$ and assume without loss of generality that $t_0=t$. This metric is a sum of independent random variables, \ie
\begin{align}\label{LambdaE}
\Lambda_\matE \triangleq\Lambda_t+\Lambda_{t+1}+\Lambda_{t+2}+\ld,
\end{align}
where $\Lambda_t=e_{1,t}l_{1,t}+e_{2,t}l_{2,t}$ corresponds to the elements defining the PEP in \eqref{PEP3}. We then express the $t$th metric as
\begin{align}\label{Lambdat}
\Lambda_t(\be_t,\bs_t) &=
\begin{cases}
0 										,& \text{if $\be_t=[0,0]$}\\
(-1)^{s_{1,t}}\tilde{l}_{1,t} 						,& \text{if $\be_t=[1,0]$}\\
(-1)^{s_{2,t}}\tilde{l}_{2,t} 						,& \text{if $\be_t=[0,1]$}\\
\sum_{k=1}^{2}(-1)^{s_{k,t}}\tilde{l}_{k,t} 			,& \text{if $\be_t=[1,1]$}\\
\end{cases},
\end{align}
where we use $\Lambda_t(\be_t,\bs_t)$ to show that $\Lambda_t$ depends on the scrambling's outcome $\bs_t$ (through $\tilde{l}_{k,t}$) and the error pattern at time $t$, $\be_t$.

Since $\tilde{l}_{k,t}$ are random variables (that depend on $k$ and $x_t$), according to \eqref{Lambdat}, there exist three pdfs that can be used to model the individual metrics in \eqref{LambdaE}. We denote the set of these three conditional pdfs by $\set{\tr{p}_{1}(\lambda|b_1),\tr{p}_{2}(\lambda|b_2),\tr{p}_{\Sigma}(\lambda|\bb)}$, for the three relevant cases defined in \eqref{Lambdat}, respectively. We note that $\tr{p}_{\Sigma}(\lambda|\bb)$ is conditioned not only on one bit, but on the pair of transmitted bits $\bb=[b_1,b_2]$, where $b_1,b_2,$ and $\bb$ represent the bits $c_{1,t}$, $c_{2,t}$, and $\bc_t$, respectively. From \eqref{LambdaE}, and due to the independence of the individual metrics, the PEP in \eqref{PEP3} can be expressed as
\begin{align}
   \tr{PEP}(\matE) 	& = \tr{PEP}_\tr{T}(w_{\matE,1}, w_{\matE,2}, w_{\matE,\Sigma})\nonumber\\
   				& = \int_{0}^{\infty} \set{\tr{p}_{1}(\lambda|b_1=0)}^{* w_{\matE,1}} * \set{\tr{p}_{2}(\lambda|b_2=0)}^{* w_{\matE,2}}* \set{\tr{p}_{\Sigma}(\lambda|\bb=[0,0])}^{* w_{\matE,\Sigma}}\,\tr{d}\lambda,\label{PEP.case3}
\end{align}
where $w_{\matE,k}$ is the number of columns in $\matE$ where only the $k$th row of $\matE$ is one, and $w_{\matE,\Sigma}$ is the number columns in $\matE$ where both entries are equal to one. Clearly
\begin{align}\label{rel_dfree}
d_\tr{H}(\matE) = w_{\matE,1}+w_{\matE,2}+2w_{\matE,\Sigma}.
\end{align}

\begin{example}[Error event at minimum Hamming distance of $(5,7)$ code]\label{Example_5_7_E}
Consider the constraint length $K=3$ optimum distance spectrum convolutional code (ODSCC) with polynomial generators $(5,7)$ \cite[Table~I]{Frenger99}. The free distance of the code is $d_\tr{H}^\tr{free}=5$, and $\beta_5^\mc{C}=1$, \ie there is one divergent path at Hamming distance five from the all-zero codeword, and the Hamming weight of that path is $d_\tr{H}(\bi_\matE)=1$. Moreover, it is possible to show that this codeword is
\begin{align*}
\matE=
\left[
\begin{array}{ccc}
1&0&1\\
1&1&1\\
\end{array}
\right],
\end{align*}
\ie $d_\tr{H}(\matE)=5$, $w_{\matE,1}=0$, $w_{\matE,2}=1$, and $w_{\matE,\Sigma}=2$. Also, $\ov{w}_{\matE,1}=2$ and $\ov{w}_{\matE,2}=3$.
\end{example}

We define $\mc{D}_{w_{1}, w_{2}, w_{\Sigma}}$ as the set of codewords $\matE$ with $w_1$ columns such that $\be_t=[1,0]$, $w_2$ columns with $\be_t=[0,1]$, and $w_{\Sigma}$ columns with $\be_t=[1,1]$, \ie $\mc{D}_{w_1,w_2,w_\Sigma}\triangleq \set{\matE\in \mc{E}:w_{1}=w_{\matE,1},w_{2}=w_{\matE,2},w_{\Sigma}=w_{\matE,\Sigma}}$. Using this, the UB expression in \eqref{UB} for BICM-T is given by
\begin{align}
\tr{UB}_\tr{T} 	&=\sum_{w_{1}, w_{2}, w_{\Sigma}} \tr{PEP}_\tr{T}(w_{1}, w_{2}, w_{\Sigma}) 
  \sum_{\matE\in\mc{D}_{w_{1}, w_{2}, w_{\Sigma}}} d_\tr{H}(\bi_\matE)\nonumber\\
 &=\sum_{w_{1}, w_{2}, w_{\Sigma}} \tr{PEP}_\tr{T}(w_{1}, w_{2}, w_{\Sigma}) \beta_{w_{1}, w_{2}, w_{\Sigma}}^\mc{C},\label{UB.case3.2}
\end{align}
where $\beta_{w_{1}, w_{2}, w_{\Sigma}}^\mc{C}$ is a weight distribution spectrum of the code $\mc{C}$ that not only considers the generalized weight $[w_1,w_2]$ of the codewords, but takes into account the temporal behavior, \ie it considers the case when $\be_t=[1,1]$ as a different kind of event. This differs from $\beta_{w_1,w_2}^\mc{C}$, where such an event will be simply considered as an extra contribution to the total generalized weight.

\subsection{PDF of the L-values}\label{Sec:Performance:PDF}

In order to calculate the PEP for BICM-T in \eqref{PEP.case3} we need the compute the set of conditional pdfs $\set{\tr{p}_{1}(\lambda|b_1),\tr{p}_{2}(\lambda|b_2),\tr{p}_{\Sigma}(\lambda|\bb)}$. In this subsection we show how to find approximations for these PDFs.

The L-values in \eqref{L-value.1} can be expressed as
\begin{align}\label{L-value.2}
\tilde{l}_{k,t}= \log\frac{\sum_{x\in\mc{X}_{k,1}}\tr{p}(y_{t}| x)}{\sum_{x\in\mc{X}_{k,0}}\tr{p}(y_{t}| x)},
\end{align}
where $\mc{X}_{k,b}$ is the set of constellation symbols labeled with $b$ at bit position $k$. Using the fact that the channel is Gaussian and if the so-called max-log approximation $\log(\tr{e}^a+\tr{e}^b)\approx \max\set{a,b}$ is used, the L-values can be expressed as
\begin{align}\label{L-value.maxlog}
\tilde{l}_{k,t}(y_{t}|\bs_t) \approx \gamma\left[\min_{x\in\mc{X}_{k,0}}(y_{t}-x)^{2} - \min_{x\in\mc{X}_{k,1}}(y_{t}-x)^{2}\right],
\end{align}
where from now on we use the notation $\tilde{l}_{k,t}(y_{t}|\bs_t)$ to emphasize that the L-values depend on the received signal and the scrambler's outcome $\bs_t$. In fact, the L-values depend on the transmitted symbol $x_t$, however, and since $\bc_t=\matzero$ and no interleaving is performed, $x_t$ is completely determined by $\bs_t$.

\begin{figure}[t]
\psfrag{Lk}[lB][lB][0.7]{$\tilde{l}_{k,t}(y_t|x_t)$}%
\psfrag{y}[cB][cB][0.7]{$y_t$}%
\psfrag{p1d}[cB][cB][0.7]{$\Delta$}%
\psfrag{p3d}[cB][cB][0.7]{$3 \Delta $}%
\psfrag{m1d}[cB][cB][0.7]{$-\Delta $}%
\psfrag{m3d}[cB][cB][0.7]{$-3 \Delta $}%
\psfrag{k1}[rB][rB][0.7]{$\tilde{l}_{1,t}(y_t|x_t)$}%
\psfrag{k2}[rB][rB][0.7]{$\tilde{l}_{2,t}(y_t|x_t)$}%
\psfrag{pL1pL2}[rB][rc][0.7]{$\tilde{l}_{\Sigma,t}(y_t|x_t,\bs=[0,0])$}%
\psfrag{mL1pL2}[rB][rc][0.7]{$\tilde{l}_{\Sigma,t}(y_t|x_t,\bs=[1,0])$}%
\psfrag{m16}[rc][rc][0.7]{$-16\gamma\Delta^2$}%
\psfrag{000}[rc][rc][0.7]{$0$}%
\psfrag{m04}[rc][rc][0.7]{$-4\gamma\Delta^2$}%
\psfrag{m08}[rc][rc][0.7]{$-8\gamma\Delta^2$}%
\psfrag{p04}[rc][rc][0.7]{$4\gamma\Delta^2$}%
\psfrag{p08}[rc][rc][0.7]{$8\gamma\Delta^2 $}%
\psfrag{p16}[rc][rc][0.7]{$16\gamma\Delta^2$}%
\psfrag{tita}[ct][ct][0.7]{a) Cases when $\be_t=[1,0]$ or $\be_t=[0,1]$}%
\psfrag{titb}[ct][ct][0.7]{b) Cases when $\be_t=[1,1]$}%
\psfrag{E10}[lc][lc][0.7]{$\be_t=[1,0]$}%
\psfrag{S0x}[lc][lc][0.7]{$\bs_t=[0,:]$}%
\psfrag{S1x}[lc][lc][0.7]{$\bs_t=[1,:]$}%
\psfrag{E01}[cc][cc][0.7][-55]{$\be_t=[0,1]$}%
\psfrag{Sx0}[cc][cc][0.7][-55]{$\bs_t=[:,0]$}%
\psfrag{Sx1}[cc][cc][0.7][55]{$\bs_t=[:,1]$}%
\psfrag{E01v2}[cc][cc][0.7][55]{$\be_t=[0,1]$}%
\psfrag{S00}[rc][rc][0.7][0]{$\bs_t=[0,0]$}%
\psfrag{S11}[rc][rc][0.7][0]{$\bs_t=[1,1]$}%
\psfrag{S10}[lc][lc][0.7][0]{$\bs_t=[1,0]$}%
\psfrag{S01}[lc][lc][0.7][0]{$\bs_t=[0,1]$}%
\psfrag{xxx}[lc][lc][0.8][0]{$y_t$}%
\psfrag{yyy}[rc][rc][0.8][0]{$\Lambda_t(\be_t,\bs_t)$}%
\begin{center}
    \includegraphics[width=0.95\columnwidth]{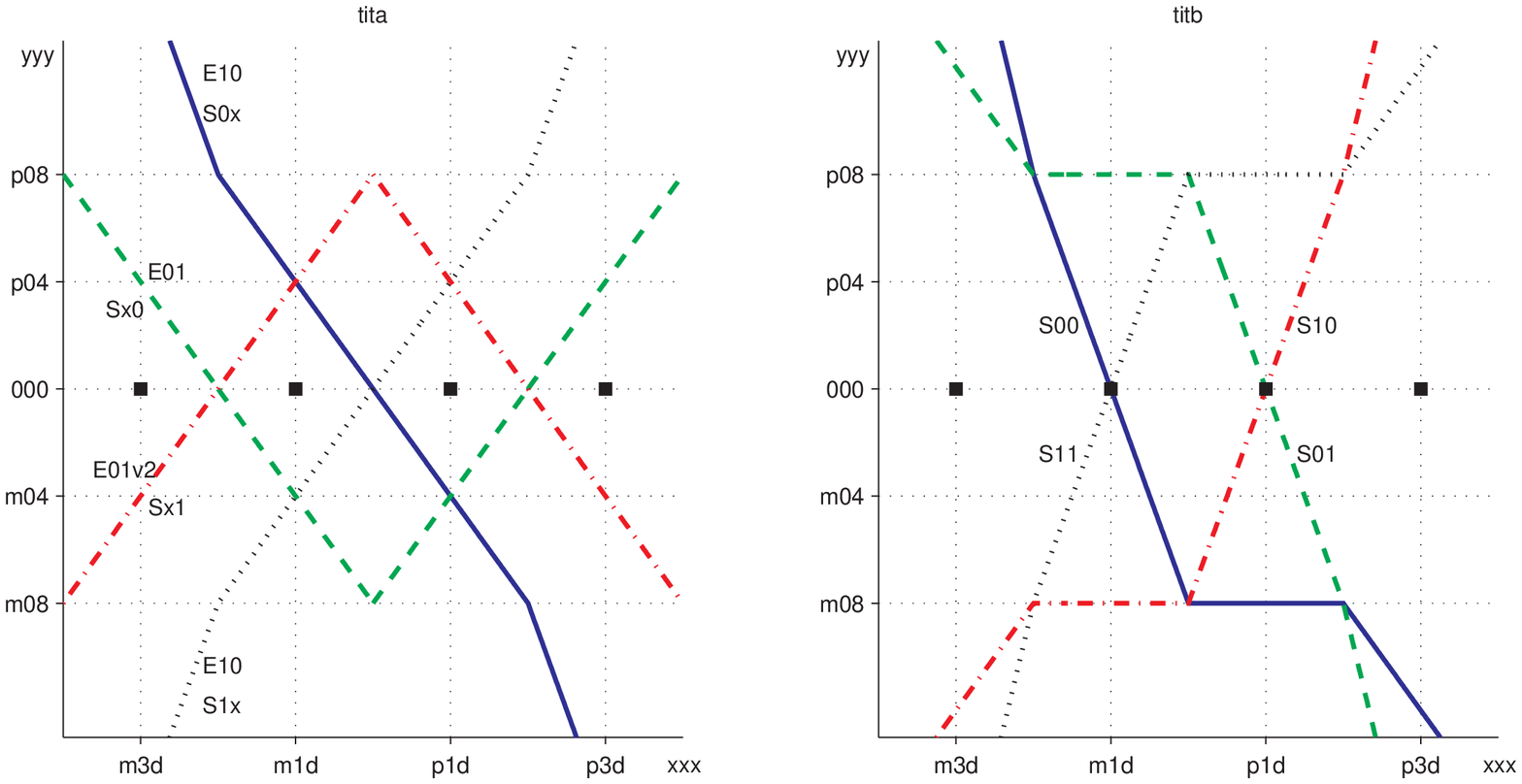}
    \caption{Piece-wise relation between the L-values $\Lambda_t(\be_t,\bs_t)$ in \eqref{Lambdat} and the received signal $y_t$ for 4-PAM for all the possible values of $\be_t$ and $\bs_t$. The relation for the case when $\be_t=[1,0]$ or $\be_t=[0,1]$ is shown in a), and the relation when $\be_t=[1,1]$ is shown in b). The transmitted symbols are shown with black squares.}
    \label{Sec:Preliminaries:PDFLLR:pcw_relation_4PAM}
\end{center}
\end{figure}

The L-value in \eqref{L-value.maxlog} is a piece-wise linear function of $y_{t}$. Moreover, the L-values $\Lambda_t(\be_t,\bs_t)$ in \eqref{Lambdat} are linear combinations of $\tilde{l}_{k,t}(y_{t}|\bs_t)$ in \eqref{L-value.maxlog}, and therefore, they are also piece-wise linear functions of $y_{t}$. Two cases are of particular interest, namely, when $\be_t=[1,0]$ or $\be_t=[0,1]$, and when $\be_t=[1,1]$. The piece-wise linear relationships for first case are shown in in Fig.~\ref{Sec:Preliminaries:PDFLLR:pcw_relation_4PAM}~a) for 4-PAM. In this figure we also show the constellation symbols and we use the notation $\bs_t=[0/1,:]$ and $\bs_t=[:,0/1]$ to show that for $\be_t=[1,0]$ and $\be_t=[0,1]$ the L-values $\Lambda_t(\be_t,\bs_t)$ are independent of $s_{2,t}$  and $s_{1,t}$, respectively. In Fig.~\ref{Sec:Preliminaries:PDFLLR:pcw_relation_4PAM}~b), the four possible cases when $\be_t=[1,1]$ are shown.

For a given transmitted symbol $x_t$ (determined by $\bs_t$), the received signal $y_t$ is a Gaussian random variable with mean $x_t$ and variance $N_0/2$. Therefore, each L-value $\Lambda_t(\be_t,\bs_t)$ in \eqref{Lambdat} is a sum of piece-wise Gaussian functions\footnote{Closed-form expressions for these pdfs of $\Lambda_t(\be_t,\bs_t)$ when $\be_t=[1,0]$ and $\be_t=[1,0]$ (\cf Fig.~\ref{Sec:Preliminaries:PDFLLR:pcw_relation_4PAM}~a) were presented in \cite{Alvarado07d}.}. In order to  obtain expressions that are easy to work with, we use the so-called zero-crossing approximation of the L-values proposed in \cite[Sec.~III-C]{Alvarado07d} which replaces all the Gaussian pieces required in the max-log model of L-values by a single Gaussian function. Intuitively, this approximation states that
\begin{align}\label{L-value.maxlog.ZC}
\Lambda_t(y_{t}|\be_t,\bs_t)\approx \hat{a}(\be_t,\bs_t)y_t+\hat{b}(\be_t,\bs_t),
\end{align}
where $\hat{a}(\be_t,\bs_t)$ and $\hat{b}(\be_t,\bs_t)$ are the slope and the free coefficient of the closest linear piece to the transmitted symbol $x_t$. 

In Table~\ref{a_b_M4} we show the values of $\hat{a}(\be_t,\bs_t)$ and $\hat{b}(\be_t,\bs_t)$ defining \eqref{L-value.maxlog.ZC} for 4-PAM, where for notation simplicity we have defined
\begin{align}\label{Alpha}
\alpha\triangleq 4\gamma\Delta^2.
\end{align}
To clarify how these coefficients are obtained, consider for example $\be_t=[0,1]$. In this case, for $\bs_t=[1,1]$, which corresponds to $x_t=-3\Delta$, the closest linear piece intersecting the $x$-axis is the left-most part of the curve labeled in Fig.~\ref{Sec:Preliminaries:PDFLLR:pcw_relation_4PAM} by $\be_t=[0,1]$ and $\bs_t=[:,1]$ (dashed-dotted line). If for example $\be_t=[0,1]$ and $\bs_t=[0,0]$ ($x_t=\Delta$), the closest linear piece is the right-most piece labeled by $\be_t=[0,1]$ and $\bs_t=[:,0]$ (dashed line). All the other values in Table~\ref{a_b_M4} can be found by a similar direct inspection of Fig.~\ref{Sec:Preliminaries:PDFLLR:pcw_relation_4PAM}.

\begin{table}[t]
\renewcommand{\arraystretch}{1.4}
\caption{Values of $\hat{a}(\be_t,\bs_t)$ and $\hat{b}(\be_t,\bs_t)$ in \eqref{L-value.maxlog.ZC} for 4-PAM found by direct inspection of Fig.~\ref{Sec:Preliminaries:PDFLLR:pcw_relation_4PAM}.}%
\label{a_b_M4}
\begin{center}
\begin{tabular}{ccccccccc}
\hline

\hline
&\multicolumn{2}{c}{$\bs_t=[1,1]$}& \multicolumn{2}{c}{$\bs_t=[1,0]$}& \multicolumn{2}{c}{$\bs_t=[0,0]$}& \multicolumn{2}{c}{$\bs_t=[0,1]$}\\
\hline

\hline
 &	$\hat{a}(\be_t,\bs_t)$&$\hat{b}(\be_t,\bs_t)$&$\hat{a}(\be_t,\bs_t)$&$\hat{b}(\be_t,\bs_t)$&$\hat{a}(\be_t,\bs_t)$&$\hat{b}(\be_t,\bs_t)$&$\hat{a}(\be_t,\bs_t)$&$\hat{b}(\be_t,\bs_t)$ \\
\hline%
$\be_t=[1,0]$ 		&$+\alpha/\Delta$&$0$&$+\alpha/\Delta $&$0$&$-\alpha/\Delta $&$0$&$-\alpha/\Delta $&$0$\\
$\be_t=[0,1]$		&$+\alpha/\Delta $&$+2\alpha$&$-\alpha/\Delta $&$-2\alpha$&$+\alpha/\Delta $&$-2\alpha$&$-\alpha/\Delta $&$+2\alpha$\\
\hline

\hline
$\be_t=[1,1]$	&$+2\alpha/\Delta$&$+2\alpha$&$+2\alpha/\Delta$&$-2\alpha$&$-2\alpha/\Delta$&$-2\alpha$&$-2\alpha/\Delta$&$+2\alpha$\\
\hline%
 
\hline
\end{tabular}
\end{center}
\end{table}

Using the approximation in \eqref{L-value.maxlog.ZC}, the L-values can be modeled as Gaussian random variables where their mean and variance depend on $\bs_t$, $\gamma$, and $\be_t$, \ie
\begin{align}\label{pLambda_s_e}
\tr{p}_{\Lambda_t}(\lambda|\be_t,\bs_t)&=\psi \bigl( \lambda; \hat{\mu}(\be_t,\bs_t), \hat{\sigma}^2(\be_t,\bs_t) \bigr),
\end{align}
where the mean value and variance are given by
\begin{align}
\label{mu}
\hat{\mu}(\be_t,\bs_t) 		& = x_t \hat{a}(\be_t,\bs_t)+\hat{b}(\be_t,\bs_t)\\
\label{sigma2}
\hat{\sigma}^2(\be_t,\bs_t) 	& = [\hat{a}(\be_t,\bs_t)]^2\frac{N_0}{2}.
\end{align}
In Table~\ref{mu_and_sigma2_M4} we show the obtained mean values and variances for the same cases presented in Table~\ref{a_b_M4}.

\begin{table}[t]
\renewcommand{\arraystretch}{1.4}
\caption{Values of $\hat{\mu}(\be_t,\bs_t)$ and $\hat{\sigma}^2(\be_t,\bs_t)$ given in \eqref{mu} and \eqref{sigma2} for 4-PAM.}%
\label{mu_and_sigma2_M4}
\begin{center}
\begin{tabular}{ccccccccc}
\hline

\hline
&\multicolumn{2}{c}{$\bs_t=[1,1]$}& \multicolumn{2}{c}{$\bs_t=[1,0]$}& \multicolumn{2}{c}{$\bs_t=[0,0]$}& \multicolumn{2}{c}{$\bs_t=[0,1]$}\\
\hline

\hline
 &	$\hat{\mu}(\be_t,\bs_t)$&$\hat{\sigma}^2(\be_t,\bs_t)$&$\hat{\mu}(\be_t,\bs_t)$&$\hat{\sigma}^2(\be_t,\bs_t)$&$\hat{\mu}(\be_t,\bs_t)$&$\hat{\sigma}^2(\be_t,\bs_t)$&$\hat{\mu}(\be_t,\bs_t)$&$\hat{\sigma}^2(\be_t,\bs_t)$ \\
\hline%
$\be_t=[1,0]$ 		&$-3\alpha$&$2\alpha$&$-\alpha$&$2\alpha$&$-\alpha$&$2\alpha$&$-3\alpha$&$2\alpha$\\
$\be_t=[0,1]$		&$-\alpha$&$2\alpha$&$-\alpha$&$2\alpha$&$-\alpha$&$2\alpha$&$-\alpha$&$2\alpha$\\
\hline

\hline
$\be_t=[1,1]$		&$-4\alpha$&$8\alpha$&$-4\alpha$&$8\alpha$	&$-4\alpha$&$8\alpha$	&$-4\alpha$&$8\alpha$\\
\hline%
 
\hline
\end{tabular}
\end{center}
\end{table}

To obtain the pdf of $\Lambda_t$ in \eqref{Lambdat}, we simply average \eqref{pLambda_s_e} over the symbols, which are assumed to be equiprobable. This results in the following expression
\begin{align}\label{Lambdat_dist}
\tr{p}_{\Lambda_t}(\lambda) &=
\begin{cases}
\frac{1}{2}\left[\psi\bigl(\lambda;-3\alpha,2\alpha\bigr)+\psi\bigl(\lambda;-\alpha,2\alpha\bigr)\right],	& \text{if $\be_t=[1,0]$}\\
\psi\bigl(\lambda;-\alpha,2\alpha\bigr),												& \text{if $\be_t=[0,1]$}\\
\psi\bigl(\lambda;-4\alpha,8\alpha\bigr),									 			& \text{if $\be_t=[1,1]$}\\
\end{cases}.
\end{align}

%Expressions \eqref{pdf_k1} and \eqref{pdf_k2} show clearly the UEP introduced by the binary labeling. For $k=2$ the L-values are modeled as a Gaussian random variable with parameters $(-\alpha,2\alpha)$. However, for $k=1$ the L-values are a Gaussian mixture where a Gaussian random variable with parameters $(-3\alpha,2\alpha)$ appears, and therefore, in average, the bit position $k=1$ offers higher protection than $k=2$.

\section{Discussion and Applications}\label{Sec:DiscAndApp}

In the previous section, we developed approximations for the pdf of the L-values passed to the decoder in BICM-T. In this section we use them to quantify the gains offered by BICM-T over BICM-S, to define asymptotically optimum CCs, and to compare BICM-T with Ungerboeck's 1D-TCM.

\subsection{Performance of BICM-T}\label{Sec:DiscAndApp:Performance}

Expression \eqref{Lambdat_dist} show the pdf of the L-values needed to compute the UB of BICM-T, \cf \eqref{PEP.case3} and \eqref{UB.case3.2}. Moreover, due to the simplifications introduced in the previous subsections the results in \eqref{Lambdat_dist} only involve Gaussian pdfs, which greatly simplifies the PEP computation in \eqref{PEP.case3}. 

\begin{theorem}\label{TheoUB}
The UB for BICM-T is
\begin{align}\label{UB.WO.final}
\tr{UB}_\tr{T} &=\sum_{w_{1}, w_{2}, w_{\Sigma}} \beta_{w_{1}, w_{2}, w_{\Sigma}}^\mc{C} \left(\frac{1}{2}\right)^{w_{1}} \sum_{j=0}^{w_{1}} \nchoosek{w_{1}}{j}Q\left(\sqrt{\frac{(w_{1}+w_{2}+4w_{\Sigma}+2j)^2}{(w_{1}+w_{2}+4w_{\Sigma})} \frac{2\gamma}{5} }\right).
\end{align}
\end{theorem}
\begin{IEEEproof}
Inserting \eqref{Lambdat_dist} in \eqref{PEP.case3}, changing the convolution of sums into a sum of convolutions, and using $\psi(\lambda;\mu_1,\sigma_1^2)*\ld * \psi(\lambda;\mu_{J},\sigma_{J}^2)=\psi(\lambda; \sum_{j=1}^{J}\mu_j, \sum_{j=1}^{J}\sigma_j^2)$, the PEP in \eqref{PEP.case3} can be expressed as
\begin{align}
 \tr{PEP}_\tr{T}(w_{1}, w_{2}, w_{\Sigma}) 	& = \int_{0}^{\infty} \left(\frac{1}{2}\right)^{w_{1}} \sum_{j=0}^{w_{1}} \nchoosek{w_{1}}{j} \psi\bigl(\lambda;\mu_{1,2,\Sigma,j},\sigma^2_{1,2,\Sigma}\bigr) \,\tr{d}\lambda,\label{PEP.final2}
 \end{align}
 where
\begin{align}\label{muj}
\mu_{1,2,\Sigma,j}					& = -(w_{1}+w_{2}+4w_{\Sigma}+2j)\alpha\\
\label{sigma_j}
\sigma^2_{1,2,\Sigma} 			& = 2(w_{1}+w_{2}+4w_{\Sigma})\alpha.
\end{align}
By using the definition of $\alpha$ in \eqref{Alpha} and $\Delta$ in \eqref{Delta}, and \eqref{muj} and \eqref{sigma_j} in \eqref{PEP.final2}, and the UB definition in \eqref{UB.case3.2}, the expression in \eqref{UB.WO.final} is obtained.
\end{IEEEproof}

\begin{figure}[t]
\psfrag{xlabel}[cB][cB][0.85]{SNR $\gamma$~[dB]}%
\psfrag{ylabel}[ct][ct][0.85]{BER}%
\psfrag{S-interleaver-looong}[cl][cl][0.9]{BICM-S \cite{Caire98}}%
\psfrag{Best-M}[cl][cl][0.9]{BICM-M (``Best'') \cite{Alvarado09c}}%
\psfrag{Best-M-WO}[cl][cl][0.9]{BICM-T (``Worst'')}%
\psfrag{Worst-M-WO}[cl][cl][0.9]{BICM-T (``Best'')}%
\psfrag{UB}[cl][cl][0.9]{UBs}%
\psfrag{AsymptoticUB}[cl][cl][0.9]{Asymptotic UB}%
\psfrag{K3}[cc][cc][0.9]{$K=3$}%
\psfrag{K8}[rc][rc][0.9]{$K=8$}%
\begin{center}
    \includegraphics[width=0.9\columnwidth]{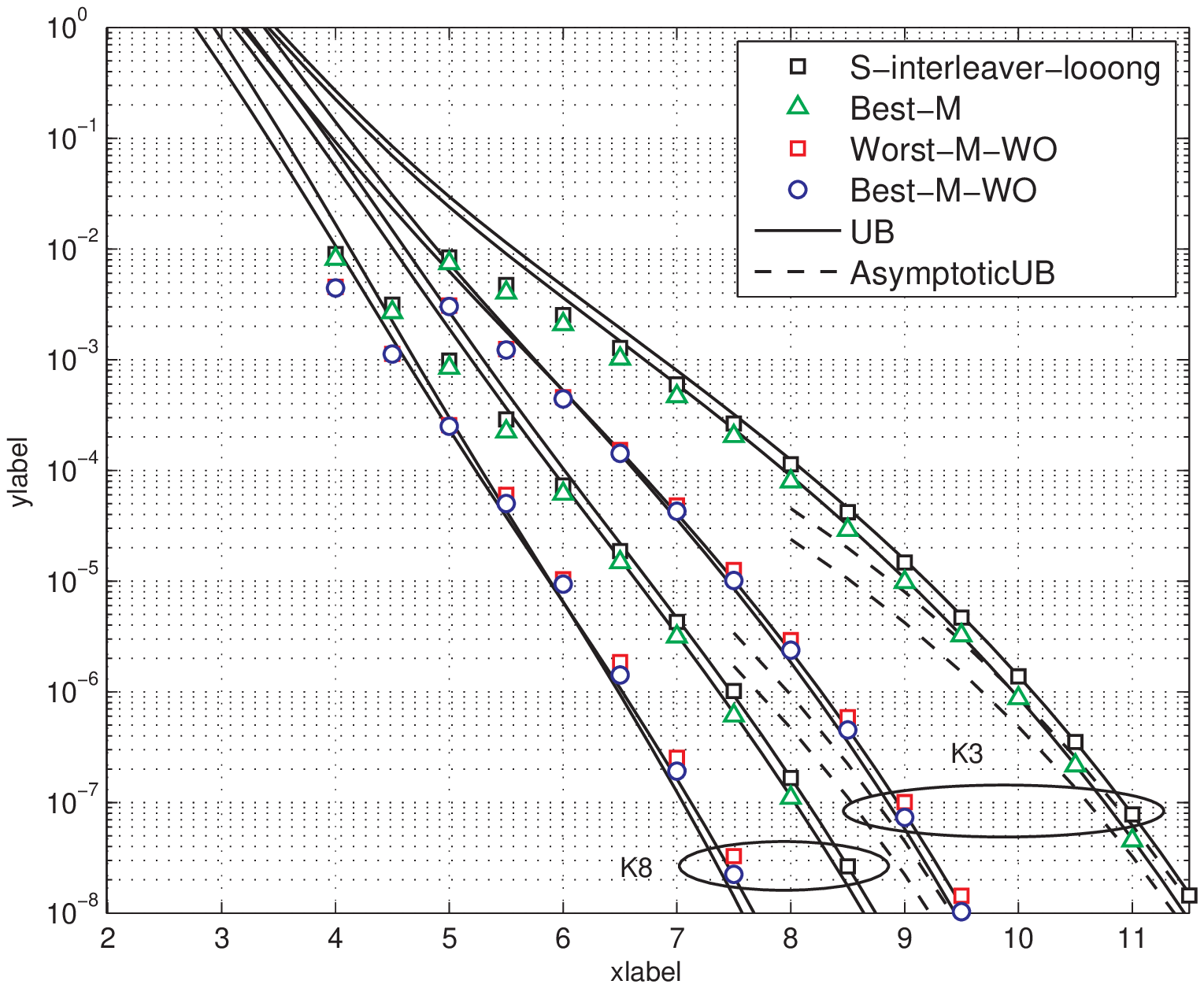}
    \caption{BER for BICM using the $(5,7)$ and $(247,371)$ ODSCCs \cite{Frenger99} and 4-PAM labeled with the BRGC, and for BICM-S \cite{Caire98}, BICM-M \cite{Alvarado09c}, and BICM-T. The simulations are shown with markers and the UB with solid lines. The asymptotic UB is shown with dashed lines.}
    \label{Sec:Performance:Numerical:n_2_m_2_K_3_different_interleavers}
\end{center}
\end{figure}

In Fig.~\ref{Sec:Performance:Numerical:n_2_m_2_K_3_different_interleavers}, numerical results for BICM-T with 4-PAM labeled with the BRGC and using the ODSCCs $(5,7)$ ($K=3$) and $(247,371)$ ($K=8$) \cite[Table~I]{Frenger99} are shown. For BICM-M two configurations are considered for each code. The first one is when all the bits from the first encoder's output are assigned to the first modulator's input and all the bits from the second encoder's output are sent to the second modulator's input. The second alternative simply corresponds to the opposite, \ie all the bits from the first encoder's output are sent over $k=2$ and the bits from the second encoder's output are sent over $k=1$. This is equivalent to defining the code by simply swapping the order of the polynomial generators. For these two particular codes, the configuration that minimizes the BER for medium to high SNR is the second one, \ie when all the bits generated by the polynomial $(7)$ or $(371)$ are sent over $k=1$ and all the bits generated by the polynomial $(5)$ or $(247)$ are sent over $k=2$. We denote the configuration that minimizes (or maximizes) the BER by ``Best'' (or ``Worst'').

To compute the UB for BICM-S and BICM-M, we use the expressions in \cite[eq.~(22)--(23)]{Alvarado09c}, and for BICM-T we use Theorem~\ref{TheoUB}. All the UB computations were carried out considering a truncated spectrum of the code, \ie $\set{w,w_1,w_2,w_\Sigma} \leq 30$ which is calculated numerically using a breadth first search algorithm \cite{Belzile93}. The results in Fig.~\ref{Sec:Performance:Numerical:n_2_m_2_K_3_different_interleavers} show that the UB developed in this paper for BICM-T predict well the simulation results. Also, these results show that for these particular codes, the gains obtained by using BICM-M instead of BICM-S are small, although larger gains were obtained in \cite{Alvarado09c} for other codes/configurations. On the other hand, the gains by using BICM-T instead of BICM-S for a BER target of $10^{-7}$ are approximately 2~dB for $K=3$ and 1~dB for $K=8$. Moreover, these gains are obtained by decreasing the complexity of the system, \ie by not doing interleaving/de-interleaving.

\subsection{Asymptotic Performance}

In this subsection, we analyze the performance of BICM-T for asymptotically high SNR and we compare it with BICM-S.

\begin{theorem}\label{asymptoicTheo}
The asymptotic UB for BICM-T and a given code $\mc{C}$ can be expressed as
\begin{align}\label{asym.UB.WO}
\tr{UB}_\tr{T}' 	&= M^\mc{C}Q\left(\sqrt{\frac{2\gamma A^\mc{C}}{5}}\right),
\end{align}
where
\begin{align}
\label{A}
A^\mc{C}	&\triangleq \min_{\substack{w_1,w_2,w_\Sigma \\ \beta_{w_{1}, w_{2}, w_{\Sigma}}^\mc{C}\neq 0}} (w_{1}+w_{2}+4w_{\Sigma})\\
\label{M}
M^\mc{C}	&= \sum_{\substack{ w_{1},w_{2},w_{\Sigma}\\ \beta_{w_{1}, w_{2}, w_{\Sigma}}^\mc{C}\neq 0 \\ w_{1}+w_{2}+4w_{\Sigma}=A}}\beta_{w_{1}, w_{2}, w_{\Sigma}}^\mc{C} \left(\frac{1}{2}\right)^{w_{1}}.
\end{align}
\end{theorem}
\begin{IEEEproof}
The UB in \eqref{UB.WO.final} is a sum of weighted Q-functions. For high SNR, and for each $(w_{1}, w_{2}, w_{\Sigma})$ there is a Q-function that dominates the the inner sum in \eqref{UB.WO.final}. This is  obtained for $j=0$, which completes the proof.
\end{IEEEproof}

For comparison purposes, we present here the performance of BICM-S at asymptotically high SNR. This can be obtained for example by particularizing \cite[eq.~(25)]{Alvarado09c} to the conventional BICM configuration with one single interleaver. The asymptotic performance of BICM-S is given by
\begin{align}\label{asym.UB.S}
\tr{UB}_\tr{S}' = \left(\frac{3}{4}\right)^{d_\tr{H}^\tr{free}}\beta_{d_\tr{H}^\tr{free}}^\mc{C} Q\left(\sqrt{\frac{2d_\tr{H}^\tr{free}\gamma}{5}}\right),
\end{align}
where $d_\tr{H}^\tr{free}$ is the free Hamming distance of the code which can be expressed as $d_\tr{H}^\tr{free}=w_{1}^\tr{free}+w_{2}^\tr{free}+2w_{\Sigma}^\tr{free}$, \cf \eqref{rel_dfree}.

In Fig.~\ref{Sec:Performance:Numerical:n_2_m_2_K_3_different_interleavers}, we show asymptotic UBs for $K=3$. For BICM-T we used Theorem~\ref{asymptoicTheo}, for BICM-S we use \eqref{asym.UB.S}, and for BICM-M we use \cite[eq.~(25)]{Alvarado09c}. 
All of them are shown to follow the simulation results quite well. Similar results can be obtained for the code with $K=8$, however, we do not show those results not to overcrowd the figure.

The asymptotic gain (AG) obtained by using BICM-T instead of BICM-S is obtained directly from Theorem~\ref{asymptoicTheo} and \eqref{asym.UB.S}, as stated in the following corollary.
\begin{corollary}\label{AG}
The AG obtained by using BICM-T instead of BICM-S
\begin{align}\label{asymgain}
\tr{AG}_{\tr{S}\rightarrow\tr{T}}	&=10\log_{10}\left(\frac{A^\mc{C}}{w_{1}^\tr{free}+w_{2}^\tr{free}+2w_{\Sigma}^\tr{free}}\right).
\end{align}
\end{corollary}

\begin{example}[AG for the $(5,7)$ code]
For the particular code $(5,7)$, it is possible to see that the solution of \eqref{A} corresponds to the event at minimum Hamming distance\footnote{However, this is not always true for other codes.}, \ie $d_\tr{H}^\tr{free}=5$, $w_{\matE,1}=0$, $w_{\matE,2}=1$, $w_{\matE,\Sigma}=2$ (\cf Example~\ref{Example_5_7_E}), and therefore, $A^\mc{C}=9$. This result in an AG of $10\log_{10}\left(\frac{9}{5}\right)\approx 2.55~\tr{dB}$. Moreover, since the input sequence that generates the codeword at minimum Hamming distance has Hamming weight one ($\beta_{0,1,2}^\mc{C}=1$), we obtain $M_{0, 1, 2}^\mc{C}=1$ for the configuration ``Worst''. If the polynomials are swapped (which corresponds to swapping the rows of $\matE$), \ie if we consider the code $(7,5)$, we obtain $w_{\matE,1}=1$, $w_{\matE,2}=0$, $w_{\matE,\Sigma}=2$  and the same $A^\mc{C}$ (since $A^\mc{C}$ does not depend on the order of the polynomials). However, in this case $M_{1, 0, 2}^\mc{C}={1}/{2}$. These two asymptotic bounds are shown in Fig.~\ref{Sec:Performance:Numerical:n_2_m_2_K_3_different_interleavers}, where the influence of the coefficient $M^\mc{C}$ can be observed in both numerical results and asymptotic bounds.
\end{example}

\subsection{Asymptotically Optimum Convolutional Codes}\label{Sec:DiscAndApp:AO}

Optimum CCs are usually defined in terms of minimum distance, \ie good CCs are the one that for a given rate and constraint length have the maximum free distance (MFD) \cite[Sec.~8.2.5]{Proakis00_Book}. The MFD criterion can be refined if the multiplicities associated to the different weights are considered \cite{Frenger99}, \cite[Sec.~12.3]{Lin03_Book}. This optimality criterion resulted in the ODSCCs which are optimal in both binary transmission and in BICM-S, \cf \eqref{asym.UB.S}. for BICM-M, we have shown in \cite{Alvarado09c} that $d_\tr{H}^\tr{free}$ is still a good indicator of the optimality of the code (as well as its multiplicity), however, a  generalized weight distribution spectrum of the code should be considered, \cf $\beta_{w_1,w_2}^\mc{C}$ in \eqref{UB.inid2}. If BICM-T is considered, and as a direct consequence of Theorem~\ref{asymptoicTheo}, asymptotically optimum convolutional codes (AOCCs) can be defined.

\begin{definition}[Asymptotically optimum convolutional codes for BICM-T]\label{AOCC-WO}
A CC is said to be an AOCC if among all codes with the same $K$ and $R=1/2$ it has the highest $A^\mc{C}$, and among all codes with the same $K$ and $R=1/2$ it has the lowest multiplicity $M^\mc{C}$.
\end{definition}

We have performed an exhaustive numerical search for AOCCs  based on Definition~\ref{AOCC-WO}. We considered  for constraint lengths $K=3,4,\ld,8$ and all  codes with free distance  $0 < d_\tr{H}^\tr{free} \leq \hat{d}_\tr{H}^\tr{free}$, where $\hat{d}_\tr{H}^\tr{free}$ is the free distance of the ODSCC. The spectrum was truncated as $w_1+w_2+4w_\Sigma\leq \hat{d}_\tr{H}^\tr{free}+8$ and the search was performed in lexicographic order. The results are shown in Table~\ref{AOCC}, where we also include the ODSCCs for comparison. If there exist more than one AOCC for a given $K$, we present the first one in the list. These results show that in general the minimum free distance of the code is not the proper criterion in BICM-T, \ie codes that are not MFD codes perform better than the ODSCCs, \cf $K=6,7$. In fact, only for $K=3$ the ODSCC is also optimum for BICM-T\footnote{For $K=4$ the AOCC $(13,17)$ has, in fact, the same spectrum $\beta_{w_1,w_2,w_\Sigma}^\mc{C}$ than the ODSCC $(15,17)$. The AOCC appears in the list because of the lexicographic order search.}. In this table we also present the AG that BICM-T offers with respect to BICM-S. The values obtained are around 2~dB.

\begin{table}[t]
\renewcommand{\arraystretch}{1.4}
\caption{AOCCs, ODSCCs, coefficients $A^\mc{C}$ and $M^\mc{C}$, and AGs.}%
\label{AOCC}
\begin{center}
\begin{tabular}{|c|cccc|cc|cc|}
\hline

\hline
\multirow{2}{*}{$K$} 
&  \multicolumn{4}{c|}{AOCCs} & \multicolumn{2}{c|}{ODSCCs} & \multicolumn{2}{c|}{AG~[dB]}\\
 		& $(g_1,g_2)$	& $d_\tr{H}^\tr{free}$ & $A^\mc{C}$		& $M^\mc{C}$	& $(g_1,g_2)$ &  $d_\tr{H}^\tr{free}$		 & $\tr{AG}_{\tr{S}\rightarrow\tr{T}}$	& $\tr{AG}_{\tr{UC}\rightarrow\tr{T}}$\\
\hline

\hline
$3$		& $(7,5)$		&5	& 9			& 0.50	& $(5,7)$		& 5			& 2.55	& 2.55	\\	
$4$		& $(13,17)$		&6	& 10			& 0.50	& $(15,17)$		& 6			& 2.22	& 3.01	\\	
$5$		& $(23,33)$		&7	& 11			& 0.38	& $(23,35)$		& 7			& 1.96	& 3.42	\\	
$6$		& $(45,55)$		&7	& 13			& 1.62	& $(53,75)$		& 8			& 2.11	& 4.15	\\	
$7$		& $(107,135)$	&9	& 14			& 0.50	& $(133,171)$	& 10			& 1.46	& 4.47	\\	
$8$		& $(313,235)$	&10	& 16			& 8.02	& $(247,371)$	& 10			& 2.04	& 5.05	\\	
\hline%
 
\hline
\end{tabular}
\end{center}
\end{table}

\begin{figure}[t]
\psfrag{xlabel1}[cc][cB][0.85]{$\mc{C}$}%
\psfrag{xlabel2}[cc][cB][0.85]{$\mc{C}$}%
\psfrag{ylabel1}[cc][ct][0.85]{$A^{\mc{C}}$}%
\psfrag{ylabel2}[cc][ct][0.85]{$M^{\mc{C}}$}%
\psfrag{ODSCC}[lc][lc][0.85]{ODSCC}%
\psfrag{AOCC}[lc][lc][0.85]{AOCC}%
\psfrag{first}[lc][lc][0.85]{$(g_1,g_2)$}%
\psfrag{second}[lc][lc][0.85]{$(g_2,g_1)$}%
\begin{center}
    \includegraphics[width=1\columnwidth]{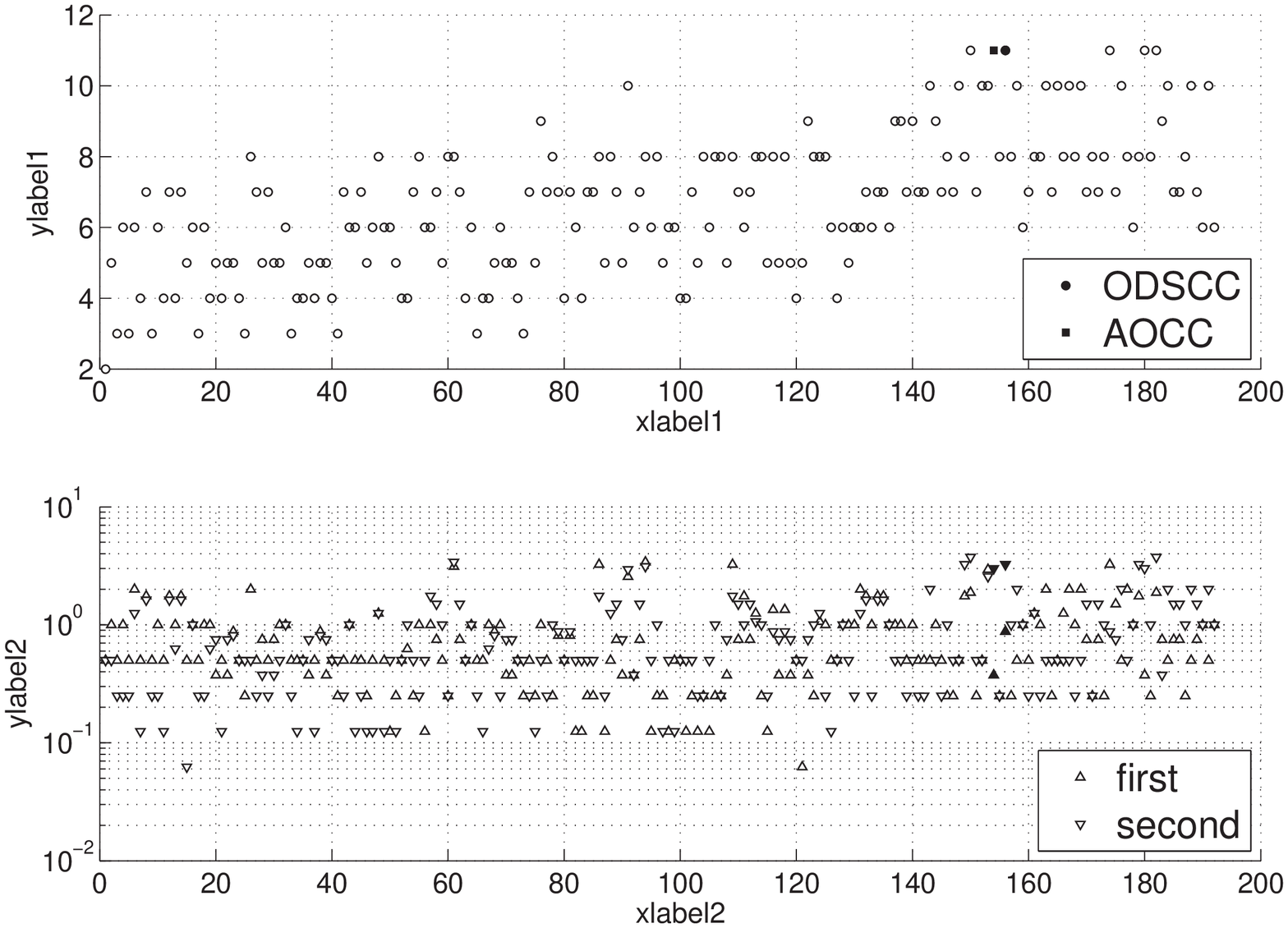}
    \caption{Values of $A^\mc{C}$ and $M^\mc{C}$ for all the possible CCs with $K=5$. The ODSCC and the AOCC are shown with filled markers.}
    \label{Sec:Performance:Numerical:AOCCs_K_5}
\end{center}
\end{figure}

\begin{figure}[t]
\psfrag{xlabel1}[cc][cB][0.85]{$\mc{C}$}%
\psfrag{xlabel2}[cc][cB][0.85]{$\mc{C}$}%
\psfrag{ylabel1}[cc][ct][0.85]{$A^{\mc{C}}$}%
\psfrag{ylabel2}[cc][ct][0.85]{$M^{\mc{C}}$}%
\psfrag{ODSCC}[lc][lc][0.85]{ODSCC}%
\psfrag{AOCC}[lc][lc][0.85]{AOCC}%
\psfrag{first}[lc][lc][0.85]{$(g_1,g_2)$}%
\psfrag{second}[lc][lc][0.85]{$(g_2,g_1)$}%
\begin{center}
    \includegraphics[width=1\columnwidth]{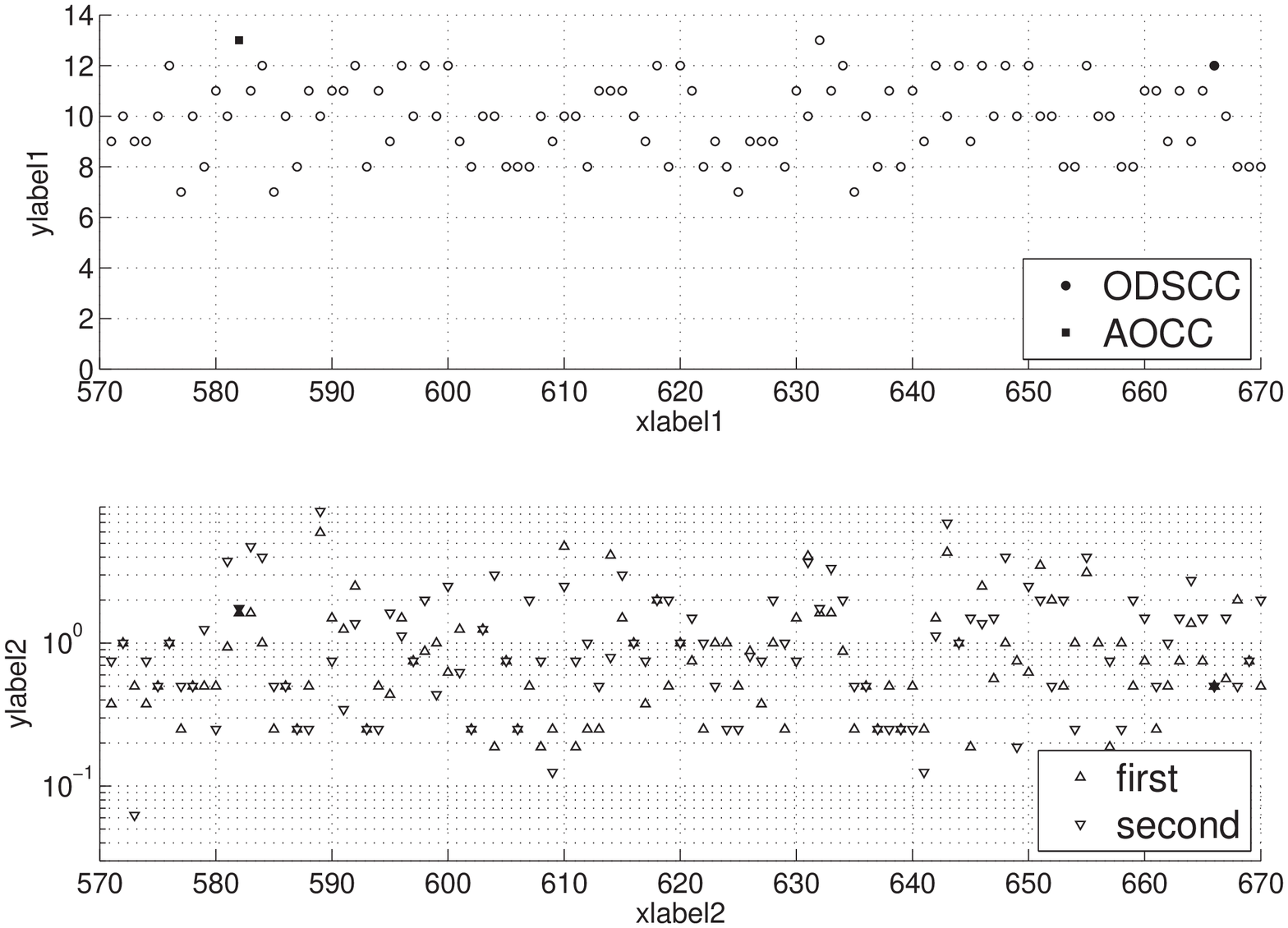}
    \caption{Values of $A^\mc{C}$ and $M^\mc{C}$ for all the possible CCs with $K=6$. The ODSCC and the AOCC are shown with filled markers.}
    \label{Sec:Performance:Numerical:AOCCs_K_6}
\end{center}
\end{figure}

In Fig.~\ref{Sec:Performance:Numerical:AOCCs_K_5}, we show the values of $A^\mc{C}$ and $M^\mc{C}$ for all the possible codes with $K=5$. This figure shows that for $K=5$ the AOCC and the ODSCCs have the same asymptotic performance (same $A^\mc{C}$), however, the multiplicity of the AOCC is smaller. In Fig.~\ref{Sec:Performance:Numerical:AOCCs_K_6}, we show similar results for $K=6$, where we only show a subset of all the possible codes. This figure shows that for $K=6$, the ODSCC gives a worse (smaller $A^\mc{C}$) asymptotic performance compared to the AOCC. Moreover, the AOCC code in this case has $d_\tr{H}^\tr{free}=7$ while the ODSCC has $d_\tr{H}^\tr{free}=8$, \cf Table~\ref{AOCC}. The same phenomenon occurs for $K=7$.

\subsection{BICM-T vs. TCM}\label{Sec:DiscAndApp:Compare}

As mentioned in Sec.~\ref{Sec:Model:Int}, the transmitter of BICM-T is identical to the transmitter of Ungerboeck's 1D-TCM. In this subsection we compare their asymptotic performance.

We have previously defined in \eqref{asymgain} the AG of BICM-T over BICM-S. It is also possible to define the AG of BICM-T compared to uncoded transmission with the same spectral efficiency (uncoded 2-PAM). Since the minimum squared Euclidean distance of the 2-PAM constellation is $4$, the AG is given by
\begin{align}\label{asymgainUC}
\tr{AG}_{\tr{UC}\rightarrow\tr{T}}	&=10\log_{10}\left(\frac{A^\mc{C}}{5}\right).
\end{align}

The AG in \eqref{asymgainUC} is tabulated in the last column of Table~\ref{AOCC}. For $K=3$, $\tr{AG}_{\tr{UC}\rightarrow\tr{T}}$ is equal to $2.55$~dB, which is the same as $\tr{AG}_{\tr{S}\rightarrow\tr{T}}$. This is because BICM-S with $K=3$ does not offer any AG compared to uncoded 2-PAM. Analyzing the results in the last column of Table~\ref{AOCC}, we find that they are the same ones obtained by 1D-TCM, \cf \cite[Table I]{Ungerboeck82}. This simply states that if BICM-T is used with the correct CC, it performs asymptotically as well as 1D-TCM, and therefore, it should be considered as good alternative for CM in nonfading channels. However, this is not the case if BICM-S is used, or if BICM-T is used with the ODSCCs.

%%%%%%%%%%%%%%%%%%%%%%%
\section{Conclusions}\label{Sec:Conclusions}

In this paper, we gave a formal explanation of why gains can be obtained when BICM-T is used in nonfading channels. BICM-T was shown to be a TCM transmitter used with a BICM receiver. An analytical model was developed and a new type of distance spectrum for the code was introduced, which is the relevant characteristic to optimize CCs for BICM-T. The analytical model was used to validate the numerical results and to show that the use of the ODSCCs, which rely on the regular minimum free distance criterion, is suboptimal.

For simplicity, the analysis presented in this paper was done only for a simple BICM configuration, and therefore, it is still unknown what the performance gains will be in a more general setup, \eg when the number of encoder's outputs is not the same as the modulator's input, for different spectral efficiencies, or when a less trivial (but still not infinitely long and random) interleaver is used. All these questions are left for further investigation.

\newpage
%%%%%%%%%%%%%%%%%%%%%%%
\bibliography{IEEEabrv,references_all}
\bibliographystyle{IEEEtran}

\end{document}